\documentclass[11pt]{article}
\usepackage{graphicx}
\usepackage{epsfig}
\usepackage[figuresright]{rotating}
\begin{document}
\begin{center}
{\bf \Large Spin polarizabilities and polarizabilities 
of the nucleon studied by
free and quasi-free Compton scattering at MAMI (Mainz)}\footnote{
Proceedings of GDH 2002, Genova, Italy 3-6 July 2002,
  M. Anghinolfi,\\ M. Battaglieri, R. De Vita, 
Eds., World Scientific 2003 (updated version)\\
Supported by Deutsche Forschungsgemeinschaft}\\[2ex]
Martin  Schumacher\\
Zweites Physikalisches Institut der Universit\"at,
Tammannstra\ss e 1 \\D-37077 G\"ottingen, Germany\\
E-mail: schumacher@physik2.uni-goettingen.de
\end{center}

\abstract{
In addition to the E2/M1 ratio of the $N \to \Delta$ transition, 
the electromagnetic polarizabilities 
and spin-polarizabilities are important 
structure constants of the nucleon which serve as sensitive tests of chiral
perturbation theory and of models of the nucleon. Recently, these quantities
have been investigated experimentally at MAMI (Mainz) by 
high-precision Compton scattering using hydrogen  and deuterium targets, where
for the latter the method of quasi-free scattering has been applied.}

\section{Introduction}
\label{sec:level1}

Compton scattering is an excellent tool for studying the electromagnetic 
structure of the nucleon. This process 
supplements  on information about  photoexcitation of  the internal degrees
of freedom of the nucleon as contained in the amplitudes for 
meson photoproduction and  supplies specific two-photon structure
constants of the nucleon as there are the electric and magnetic
polarizabilities $\alpha$ and $\beta$, respectively, and the
spin-polarizabilities $\gamma_0$ and $\gamma_\pi$ for the forward
and backward directions, respectively. These two-photon structure
constants are special cases of the invariant
(LPS) \cite{lvov97} amplitudes $A_i(\nu,t)\,\,(i=1 \cdots 6)$ which, therefore,
may be considered as generalized polarizabilities. One motivation for
investigating the generalized polarizabilities  $A_i$ 
is that insight is obtained into the physical nature of the two-photon
structure constants. Specifically, t-channel phenomena may be investigated
as there are the $\sigma$ meson contribution to $\alpha - \beta$ and the
$\pi^0$, $\eta$
and $\eta'$  meson contributions to  $\gamma_\pi$. 
While the t-channel part of $\gamma_\pi$ appears
to be well
understood, the corresponding part  of   $\alpha - \beta$ is still
under
investigation and especially awaits consideration  in chiral
perturbation theory. 
Recently, deuteron  targets have  been  used in measurements
of quasi-free Compton scattering by the nucleons 
\cite{kossert02,levchuk94} bound in the deuteron,  thus leading to 
the first precise results for the polarizabilities of the neutron.
In contrast to the competing  methods of electromagnetic scattering
of neutrons in a Coulomb field and coherent elastic (Compton) scattering
of photons by the deuteron where model uncertainties cannot be excluded,
the method of  quasi-free scattering has been tested and found valid
through free and quasi-free Compton scattering experiments on the proton.

\section{Polarizabilities and invariant amplitudes}

In  the extreme forward ($\theta=0$) and extreme backward
directions ($\theta=\pi$) the amplitudes for Compton scattering
may be written in the form
\cite{lvov97,babusci98}  
\begin{eqnarray}
T^{\rm LAB}(\theta=0)&=&f_0(\omega){\epsilon}'\cdot{\epsilon}+
g_0(\omega)\, \mbox{i}\, {\sigma}\cdot({\epsilon}'\times
{\epsilon})\nonumber\\
T^{\rm LAB}(\theta=\pi)&=&f_\pi(\omega){\epsilon}'\cdot{\epsilon}+
g_\pi(\omega)\, \mbox{i}\,{\sigma}\cdot({{\epsilon}}'
\times
{{\epsilon}}).
\label{T2}
\end{eqnarray}
where
$f(\omega)= 1/2 \{T_{1/2}+T_{3/2}\}$ 
correspond to the case where the initial-state and final-state photons 
have parallel linear polarization and 
$g(\omega)= 1/2 \{T_{1/2}-T_{3/2}\}$ to the case where these 
photons have perpendicular linear polarization. The relations between 
the amplitudes $f$ and $g$ and the invariant (LPS) \cite{lvov97}
amplitudes $A_i$ are \cite{babusci98}
\begin{eqnarray}
&&f_0(\omega)= -\frac{\omega^2}{2\pi}\left[A_3(\nu,t)+ A_6(\nu,t)
\right],\nonumber\\
&&g_0(\omega)=\frac{\omega^3}{2\pi m}A_4(\nu,t), \nonumber\\
&&f_\pi(\omega)=-\frac{\omega \omega'}{2\pi}\left(1+\frac{\omega\omega'}
{m^2}\right)^{1/2}\left[ 
A_1(\nu,t) + \frac{\omega\omega'}{m^2}A_5(\nu,t)\right],\nonumber\\
&&g_\pi(\omega)=-\frac{\omega \omega'}{2\pi}\left(1+\frac{\omega\omega'}
{m^2}\right)^{-1/2}\frac{\omega+\omega'}{2m}\left[
A_2(\nu,t)+ \left(1+\frac{\omega\omega'}{m^2}\right)A_5(\nu,t)\right]
\nonumber,\\
&&\omega'(\theta=\pi)=\frac{\omega}{1+2\frac{\omega}{m}},\,
\nu=\frac12 (\omega+\omega'),\, t(\theta=0)=0,
\,t(\theta=\pi)=-4\omega\omega.'
\label{T3}
\end{eqnarray}
For the polarizabilities we obtain the relations
\begin{eqnarray}
&&\alpha+\beta = -\frac{1}{2\pi}\left[A^{\rm nB}_3(0,0)+ 
A^{\rm nB}_6(0,0)\right], \,
\alpha-\beta = -\frac{1}{2\pi}
\left[A^{\rm nB}_1(0,0)\right], \nonumber\\ 
&&\gamma_0= \frac{1}{2\pi m}\left[A^{\rm nB}_4(0,0)
\right], \,
\gamma_\pi = -\frac{1}{2\pi m}
\left[A^{\rm nB}_2(0,0)+A^{\rm nB}_5(0,0) \right], 
\label{T4}
\end{eqnarray}
where $A^{\rm nB}_i$ are the non-Born parts of the amplitudes $A_i$.

Measurements of spin independent and spin dependent total photoabsorption 
cross sections have  been carried out. 
Using the sum rules
\begin{equation}
\alpha+\beta=\frac{1}{2\pi^2}\int^\infty_{\omega_{thr}} 
\sigma_{tot}(\omega)
\frac{d\omega}{\omega^2} \quad   \mbox{and} \quad
\gamma_0=\frac{1}{4 \pi^2}\int^\infty_{\omega_{thr}}
\frac{\sigma_{1/2}(\omega)-\sigma_{3/2}(\omega)}{\omega^3}d\omega,
\label{baldin}
\end{equation}
respectively,
the following results have been obtained \cite{levchuk00,ahrens00,ahrens01}
\begin{equation}
\alpha_p+\beta_p= 14.0\pm 0.3, \quad \alpha_n+\beta_n= 15.2\pm 0.5,
\quad \gamma^{(p)}_0 =  -0.86 \pm 0.13  
\label{alphabetagamma}
\end{equation}
in units of $10^{-4}{\rm fm}^3$ for polarizabilities and $10^{-4}{\rm fm}^4$
for the spin polarizabilities. The number given for the spin 
polarizability $\gamma^{(p)}_0$
contains sizable corrections for low and high energy contributions
not covered by the experiment,
the errors of which have not been taken into account.

\section{Asymptotic amplitudes}

According to the Landau-Yang theorem \cite{yang50} particles with 
$J^{PC}=0^{++},2^{++},4^{++},\cdots$ decay into  two photons with
parallel directions of linear polarization and  particles with 
$J^{PC}=0^{-+},2^{-+},4^{-+},\cdots$ into two photons with
perpendicular  directions of linear polarization for the case of total
helicity $\lambda= \lambda_{\gamma_2}-\lambda_{\gamma_1}=0$.
This  leads  to the conclusion that mainly the $\sigma(600)$  meson 
may be a $t-$channel contribution
to $f_\pi(\omega)$, whereas mainly the 
$\pi(135)$, $\eta(547)$ and $\eta'(958)$ 
mesons may be $t-$channel contributions to $g_\pi(\omega)$. 
For the case of total helicity 
$\lambda= \lambda_{\gamma_2}-\lambda_{\gamma_1}=2$ the corresponding 
quantum numbers are $J^{PC}=2^{++},4^{++},6^{++},\cdots$
and $J^{PC}=3^{++},5^{++},7^{++},\cdots$. From this it may be
concluded that 
the Pomeron ${\mathcal P}$, the $f_2(1270)$ and the 
$a_2(1320)$ are   $t-$channel contributions to $f_0(\omega)$
which are used to parametrize the total 
photoabsorption cross section based on  a Regge ansatz and, thus, 
are already taken care of by the $s-$channel. 
A natural extension of these considerations is
that $J^{PC}=3^{++}$  mesonic or gluonic intermediate states
may be  $t-$channel contributions to  the Gerasimov-Drell-Hearn 
\cite{gerasimov66} amplitude
$g_0(\omega)$. 

\section{Results of Compton scattering by the proton and the neutron}
The four polarizabilities (\ref{T4}) are defined for the forward 
($\theta=0$) and backward ($\theta=\pi$) directions and for low energies. 
Experiments,
however, require intermediate angles and energies where the terms of
higher order in $\omega$ are not negligible. Therefore, the
Compton scattering process has to be investigated in general. 
In the second resonance region and at backward angles
the data mainly serve  as a test of the 
$\sigma$-pole ansatz \cite{lvov97} for the asymptotic part of $f_\pi$
and for the determination of the relevant mass-parameter  $m_\sigma$
of this ansatz. A good fit to the data of the second resonance region is 
obtained \cite{galler01,wolf01} if a parameter of $m_\sigma=600$ MeV is 
applied. This result appears to be quite satisfactory because 
this number is in agreement with results of other investigations. 
Nevertheless, more precise data extending to higher energies 
and disentangling the amplitudes $f_\pi$ and $g_\pi$ via polarized
photons  would be desirable. This makes Compton scattering in the second 
resonance region an ideal project for the  polarized photon
beams at the 1.5 GeV accelerator MAMI C (Mainz) and at 
GRAAL (Grenoble).

Compton scattering by the proton at energies below $\pi$ threshold
using the large-angle arrangement TAPS led to a determination
of $\alpha_p -\beta_p$ with unprecedented precision \cite{olmos01}.
The result obtained
\begin{equation}
\alpha_p-\beta_p= 10.5\pm 0.9({\rm stat + syst})\pm 0.7({\rm model})
\label{alpha-beta}
\end{equation}
serves as a standard input for all analyses of Compton scattering
by the proton in terms of dispersion theories. 
 
Precise  results for the electromagnetic polarizabilities of the neutron 
have been
obtained for the first time in a recent  experiment on 
quasi-free Compton scattering by the
neutron carried out \cite{kossert02} using the large Mainz
48 cm $\oslash$ $\times$ 64 cm  NaI(Tl) detector and the G\"ottingen
segmented neutron detector SENECA in coincidence. 
Furthermore, quasi-free Compton scattering by the proton 
and Compton scattering by the free proton have been analyzed 
as tests of the method.
As a byproduct the experiment carried out for the free proton served 
to determine the spin polarizability $\gamma^{(p)}_\pi$ leading to
\cite{camen02}
\begin{equation}
\gamma^{(p)}_\pi=(-38.7 \pm 1.8).
\label{gammapi}
\end{equation}
The essential conclusion drawn from this result is that the spin
polarizability $\gamma^{(p)}_\pi$ is in agreement with the prediction
of dispersion theory (\ref{T4}) and not in strong disagreement as 
observed in \cite{tonnison98}. Therefore, for the  analysis
of the quasi-free double differential cross sections for the
neutron shown in the left panel of Fig. 1 we also may assume,
that $\gamma^{(n)}_\pi$ is in agreement with the prediction of dispersion
theory, leading to the value  given in the figure caption. 
The fit to the double differential
cross sections then led to the result 
\begin{equation}
\alpha_n - \beta_n= 9.8 \pm 3.6({\rm stat}){}^{+2.1}_{-1.1}({\rm syst})
\pm 2.2({\rm model}).
\label{alpha-beta-n}
\end{equation}
It is of interest to note that the number adopted for $\gamma^{(n)}_\pi$
is confirmed by the experiment as is shown in the right panel of Fig. 1
where the two parameter $\alpha_n - \beta_n$ and $\gamma^{(n)}_\pi$
are varied independently. Making use of the $\chi^2$ distribution we arrive 
at 
\begin{equation}
\gamma^{(n)}_\pi=58.6 \pm 4.0.
\label{gammapi-2}
\end{equation}
\begin{figure}
\centerline{\psfig{file=qf_n.eps,width=6cm,clip=,silent=,angle=0}\hfill
            \psfig{file=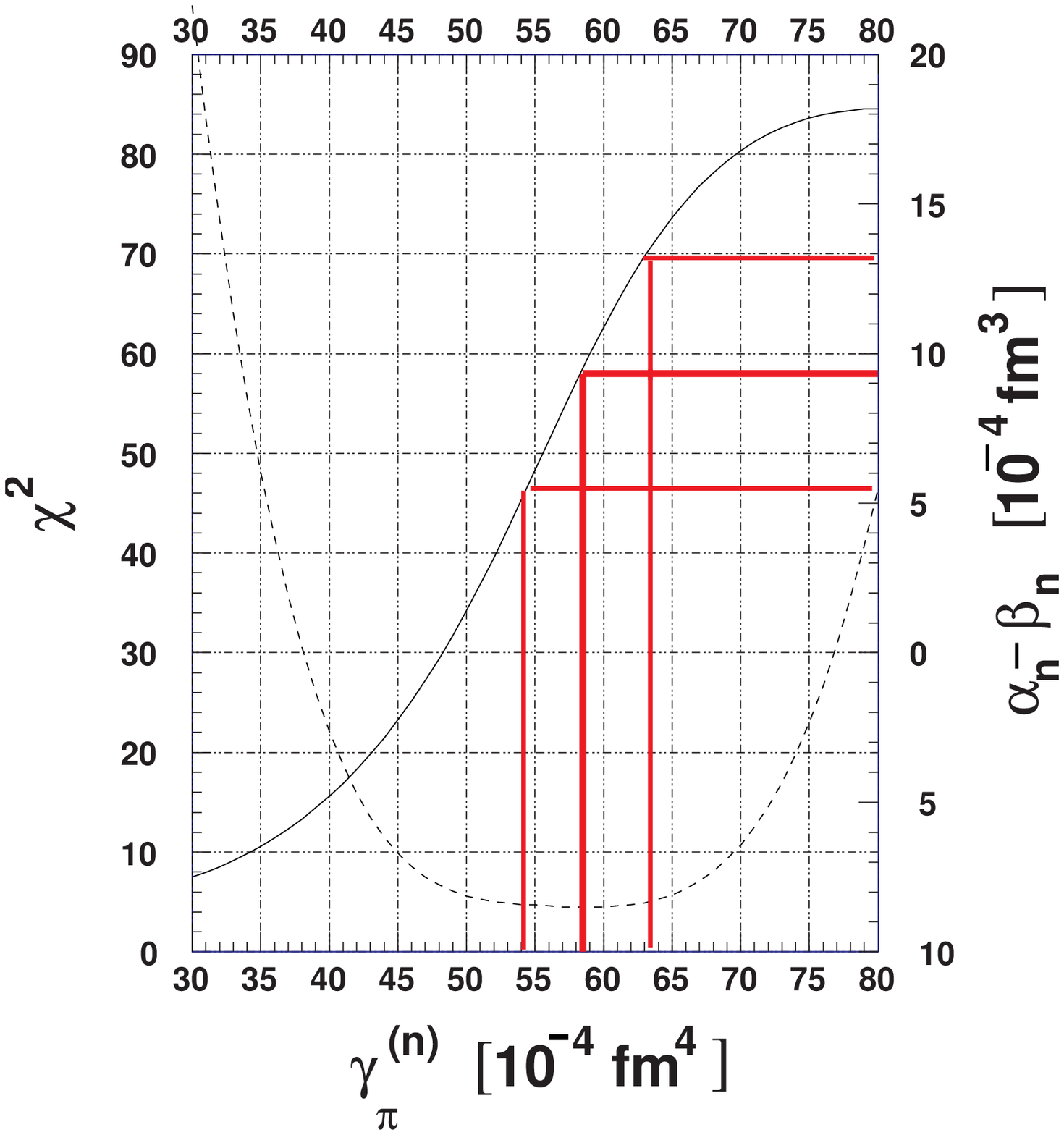,width=6cm,clip=,silent=,angle=0}}
\caption{left: Triple differential cross section for 
Compton scattering by the neutron compared with prediction
using $\alpha_n - \beta_n=
9.8 \times 10^{-4}{\rm fm}^3$ and $\gamma^{(n)}_\pi=58.6 \times
10^{-4}{\rm fm}^4$.
right: $\chi^2$ and $\alpha_n-\beta_n$ versus 
$\gamma^{(n)}_\pi$}
\label{fig:subfigures}
\end{figure}

A summary of the status of electromagnetic polarizabilities
is given in the followig table. The experimental data 
for $\alpha$, $\beta$ and $\gamma_\pi$ are due to
recent experimental work carried out at MAMI using  LARA
\cite{galler01,wolf01},  TAPS \cite{olmos01} and
the large Mainz NAI(Tl) combined with the SENECA detector
\cite{kossert02,camen02}. The value for
$\gamma_\pi(t-{\rm channel})$ is based on the prediction provided by 
(\ref{e10}),
the two values  for $\gamma_\pi(s-{\rm channel})$ labeled 
"present work" and "sum rule" on (\ref{e11}) and (\ref{e12}), respectively:
\begin{table}[t]
\label{table}
\caption{Summary of polarizabilities of proton and neutron}
\begin{center}
\begin{tabular}{llll}
\hline
& proton&neutron&\\
\hline
$\alpha$ & 12.2 $\pm$ 0.6& 12.5 $\pm$ 2.3& experiment\\ 
$\beta$ & 1.8 $\mp$ 0.6& 2.7 $\mp$ 2.3& experiment\\
$\gamma_\pi$& -38.7$\pm$1.8&+58.6$\pm$4.0& experiment\\
$\gamma_\pi$& -39.5$\pm$2.4&+52.5$\pm$2.4&sum rule \cite{lvov99}\\
\hline
$\gamma_\pi({\rm t-channel})$& -46.6 &+43.4&$\pi^0+\eta+\eta'$\\
$\gamma_\pi({\rm s-channel})$& +7.9 $\pm$ 1.8& +15.2$\pm$ 4.0&
present work\\
$\gamma_\pi({\rm s-channel})$& +7.1 $\pm$ 1.8& +9.1$\pm$ 1.8&
sum rule \cite{lvov99}\\
\hline
\end{tabular}
\end{center}
\end{table}
\begin{eqnarray}
&&\gamma_\pi(t-{\rm channel})
=\frac{1}{2\pi m}
\left[ \frac{g_{\pi NN}F_{\pi^0 \gamma\gamma}}{m^2_{\pi^0}}\tau_3
+\frac{g_{\eta  NN}F_{\eta \gamma\gamma}}{m^2_{\eta}}
+\frac{g_{\eta'  NN}F_{\eta' \gamma\gamma}}{m^2_{\eta'}} \right],
\label{e10}\\
&&\gamma_\pi(s-{\rm channel})=- \, \frac{1}{2\pi m}\left[A^{\rm int}_2(0,0)
+  A^{\rm int}_5(0,0) \right], \hspace{5cm} \label{e11}\\
&&\gamma_\pi(s-{\rm channel})=\int^\infty_{\omega_0}\sqrt{1+\frac{2\omega}{m}}
\left( 1+\frac{\omega}{m}\right) 
 \sum_nP_n \left[ \sigma^n_{3/2}(\omega) -  \sigma^n_{1/2}(\omega) \right]
\frac{d\omega}{4\pi^2 \omega^3},\label{e12}
\end{eqnarray}
where $A^{\rm int}_i$ denotes the integral part of the amplitude and 
$P_n=\pm 1$  the relative parity of the final state $n$ with
respect to the target \cite{lvov99}.

There is good agreement between the present work and the sum
rule prediction \cite{lvov99}
for  $\gamma_\pi(s-{\rm channel})$ in case of the proton
but some discrepancy between the corresponding numbers obtained for the 
neutron. We believe that this discrepancy is  caused by internal
inconsistencies of the photomeson amplitudes available for the neutron.

\end{document}